\documentclass[%
preprint,
 amsmath,amssymb,
 aps,
]{revtex4-2}

\usepackage{graphicx}
\usepackage{dcolumn}
\usepackage{bm}

\newcommand{\ag}[1]{{#1}^1A_g}

\newcommand{\agp}[1]{{#1}^1A_g^+}
\newcommand{\bup}[1]{{#1}^1B_u^+}
\newcommand{\agm}[1]{{#1}^1A_g^-}
\newcommand{\bum}[1]{{#1}^1B_u^-}

\begin{document}

\preprint{APS/123-QED}

\title{Triplet-Pair Character of the $2^1A_g$ Dark State of Polyenes}

\author{Alexandru G. Ichert}
\affiliation{Department of Chemistry, Physical and Theoretical Chemistry Laboratory, University of Oxford, Oxford, OX1 3QZ, UK}
\affiliation{Linacre College, University of Oxford, Oxford, OX1 3JA, UK}
\email{alexandru.ichert@chem.ox.ac.uk}

\author{William Barford}
\affiliation{Department of Chemistry, Physical and Theoretical Chemistry Laboratory, University of Oxford, Oxford, OX1 3QZ, UK}
\affiliation{Balliol College, University of Oxford, Oxford, OX1 3BJ, UK}
\email{william.barford@chem.ox.ac.uk}

\date{\today}

\begin{abstract}
We define and calculate the triplet-pair population of the $2^1A_g$ dark state in polyenes, as predicted by the $\pi$-electron Pariser-Parr-Pople (PPP) model, for chains of 8 to 14 C-atoms and Coulomb interaction parameter between 4-14 eV. Our definition of the triplet-pair population  is motivated by a two-particle model of the $2^1A_g$ state. We use DMRG to solve the PPP model and we exploit the MPS representation of the DMRG wavefunction to compute the triplet-pair population.
Using our results for short chain sizes, we predict a finite-size scaling value of the triplet-pair population of ca.\ 75\% for realistic Coulomb interactions for polyene chains. Our results agree with other theoretical work on the doubly-excited character of polyenes, and represents further evidence that the $\ag{2}$ state is predominantly triplet-pair in character  - with implications for singlet fission mechanisms in polyenes. 
\end{abstract}

\maketitle

\section{Introduction}

The $2^1A_g$ dark state in polyenes has been of interest to researchers for over 50 years~\cite{1972Schulten,1972Hudson}. There is general agreement to its mixed triplet-pair and charge-transfer exciton character in linear $\pi$-conjugated systems~\cite{1972Schulten,1999Chandross,WB_Book,2022Barford, 2020Valentine}, although limited effort has been directed towards quantifying the doubly excited character~\cite{2023Plasser,2025Plasser,2022Barford, 2016Dreuw, 2019Boguslawski}.

In 1972, Schhulten and Karplus were the first to predict the triplet-pair (or bimagnon) character of the $2^1A_g$ state~\cite{1972Schulten}; they showed that inclusion of doubly-excited determinants reproduced the relative energy ordering of the dark and bright states observed experimentally~\cite{1972Hudson}. In 1976, Schulten et al.~\cite{1976Schulten} further investigated the nature of the dark state and identified that double excitations make the dominant contribution to this wavefunction. Ten years later, Hayden and Mele compared the relaxed lattice configurations for the $2^1A_g$ state with that of two $1^3B_u$ states and noted a striking resemblance between the two for a chain of 16 C-atoms~\cite{1986Hayden}. Further significant theoretical contributions were made in 1987 by Tavan and Schulten~\cite{1987Tavan}, who showed that covalent excitations are combinations of triplet excitations in long polyenes. Importantly, the lowest covalent state, namely the $2^1A_g$ state, has a bond structure described by a pair of neutral soliton-antisoliton pairs, or equivalently a triplet-pair (where each triplet has a neutral soliton-antisoliton structure). Additional evidence for the four-soliton structure of the dark state in long polyenes was provided by Barford et al.~\cite{1999Barford, 2001Barford} for chains of up to 50 ethylene dimers.

Chandross et al.~\cite{1999Chandross} used a diagrammatic exciton-basis valence bond description of polyenes to investigate the low-lying eigenstates of short chains. Their work showed that the $2^1A_g$ state is best referred to as a mixture of triplet-pair and charge-transfer exciton components. Likewise, Valentine et al.~\cite{2020Valentine} calculated the exciton wavefunction of the $2^1A_g$ state and showed that it corresponds to the lowest member of a family of electron-hole excitations, which they named the ``$2A_g$-family'' of states. They concluded that the dark state is a linear combination of odd-parity charge-transfer excitons and $T_1 \otimes T_1$ (i.e., $1^3B_u \otimes 1^3B_u$) triplet-pairs. The hybridization between the exciton and triplet-pair components was investigated in more detail by Barford~\cite{2022Barford}, who showed that the coupling between the two subspaces, mediated by a one-electron transfer between neighboring triplets, was responsible for the strong binding of the triplet-pair.

More recently, do Casal et al.~\cite{2023Plasser} used a definition of excited state character based on the one-electron transition density matrices to show that the optically allowed $1^1B_u$ state is almost exclusively singly-excited in character, whereas the $2^1A_g$ state exhibits large multiply-excited contributions with respect to the ground state. Chagas et al.~\cite{2025Plasser} performed a similar calculation on polyenes of 6, 8, 10 and 12 C-atoms and observed that the doubly-excited character of the $\ag{2}$ state was above 50\% in all cases, which increases with increasing chain size. This result complements the previous findings of Starcke et al.~\cite{2016Dreuw}, who calculated the population of doubly-excited configurations in the $2^1A_g$ state of chains of two to six ethylene dimers to be ca.\ 80\%. Furthermore, similar results were obtained by Boguslawski~\cite{2019Boguslawski} who used a pair coupled-cluster double (pCCD) method to calculate the double excitation contribution to the $2^1A_g$ wavefunction for short chains.

In addition to  the purely theoretical interest in understanding the nature of the dark state, there is experimental~\cite{Kraabel98,Lanzani99,Tauber2010,Musser13,Musser19,Dasgupta2021} and theoretical~\cite{Zimmerman2018,2020Valentine,2022Santra,2023Chambers,2023Barford,2025Ichert,2025Santra} evidence which suggests that the $2A_g$-family of states could act as intermediates in the formation of uncorrelated triplets during singlet fission in polyenes and carotenoids. In long polyene-type systems there is evidence of triplet formation via singlet fission on single chains~\cite{Kraabel98,Lanzani99,Musser13}. For shorter polyenes, e.g., carotenoids, however, singlet fission appears to occur only in aggregates and dimers~\cite{Zhu2014,Musser15,Llansola-Portoles2018,Dasgupta2021,Quaranta2021,Peng2024}.

There is some controversy as to whether or not singlet fission in carotenoid aggregates proceeds via an intramolecular singlet triplet-pair intermediate.
On one hand, Kundu and Dasgupta~\cite{Dasgupta2021} and Quaranta et al.~\cite{Quaranta2021} argue that singlet fission in lycopene H-aggregates does proceed via such a state. 
Conversely, however, as for polyacenes, Musser et al.~\cite{Musser15}, Aryanpour et al.~\cite{Aryanpour2015}, and  Peng et al.~\cite{Peng2024} argue that singlet fission in  carotenoid dimers and aggregates proceeds from the optically bright state to a bimolecular triplet-pair state via a bimolecular charge-transfer exciton.

To help elucidate the singlet fission mechanism in polyenes and carotenoids, it is necessary to quantify the triplet-pair character of the dark states. In this paper we achieve this goal by solving the Pariser-Parr-Pople (PPP) model of $\pi$-electron systems using the density matrix renormalisation group (DMRG) method~\cite{1992White,1993White,DMRGSchollwock, MPSSchollwock}. The triplet-pair population of the $\ag{2}$ state as a function of the Coulomb interaction is computed by exploiting the MPS structure~\cite{1995Ostlund} of the DMRG wavefunction. Our rigorous DMRG solution of the PPP model means that the highly-correlated $\ag{2}$ state is correctly described. Moreover, our real-space projection technique computes a triplet-pair population that is relevant to the expected triplet yield from singlet fission.

The contents of this paper are as follows. Section~\ref{Se:2} contains the theoretical details. We begin  by describing the PPP model (in Section \ref{Se:2.1}), and reprising the DMRG method and MPS formulation (in Section \ref{Se:2.2}). Next, in Section \ref{Se:2.3} we motivate our definition of the triplet-pair population of the  $\ag{2}$ state by describing an exact calculation of the two-particle population of a toy model that describes two-particle interactions. This method is adapted in Section \ref{Se:2.4} to calculate the triplet-pair population of the  $\ag{2}$ state computed using DMRG from the PPP model. Our results are described in Section \ref{Se:3} and we conclude in Section \ref{Se:4}.


\vfill
\pagebreak

\section{Theory}\label{Se:2}

\subsection{The Pariser-Parr-Pople model}\label{Se:2.1}

The Pariser-Parr-Pople (or extended Hubbard) Hamiltonian for a linear chain of $N$ C-atoms is defined as
\begin{widetext}
\begin{eqnarray}\label{eq:PPP_Ham}
    \hat{H}_{PPP}= && -\sum_{i \sigma}^{N-1}t_i \left( \hat{c}_{i\sigma}^\dagger \hat{c}_{i+1\sigma} + \hat{c}_{i+1\sigma}^\dagger \hat{c}_{i\sigma} \right)
    + U\sum_{i} \left( \hat{N}_{i\uparrow}-\frac{1}{2} \right)\left( \hat{N}_{i\downarrow}-\frac{1}{2} \right)\\ \nonumber
   && +\sum_i\sum_{j\geq i}V_{ij} \left( \hat{N}_i-1 \right)\left( \hat{N}_{j}-1 \right).
\end{eqnarray}
\end{widetext}
The operator $\hat{c}_{i\sigma}^\dagger$ ($\hat{c}_{i\sigma}$) creates (destroys) an electron with spin $\sigma$ in the $p_z$ orbital on C-atom $i$. The hopping of the electrons between neighboring atoms is  defined by the transfer integral $t_i$, where for periodic bond alternation we define $t_d = t_0(1+\delta)$ and $t_s = t_0(1-\delta)$ as the double- and single-bond hopping integrals, respectively. The Coulomb interaction is modeled using the semi-empirical Ohno potential,
\begin{equation}
    V_{ij} = \frac{U}{\sqrt{1 + \left( U \epsilon r_{ij}/14.397 \right)^2}},
\end{equation}
where the Coulomb parameter $U$ is in eV, $\epsilon$ is the relative permittivity and $r_{ij}$, the distance between electrons on C-atoms $i$ and $j$, is in $\textrm{\AA}$. We vary $U$ between 4-14 eV, range which includes the typical values of the Coulomb parameter used for $\pi$-conjugated systems~\cite{1997Chandross,2020Valentine}. For all calculations that we discuss next, we have taken $\epsilon = 2$, $t_0 = 2.4$ eV and $\delta = 1/12$, so that $t_d = 2.6$ eV and $t_s = 2.2$ eV. 

\subsection{DMRG and the MPS representation}\label{Se:2.2}

We use the Density Matrix Renormalisation Group (DMRG) method~\cite{1992White,1993White, DMRGSchollwock} to solve the PPP Hamiltonian. The DMRG algorithm accurately and efficiently truncates the Hilbert space, and is especially well-suited at solving one-dimensional problems~\cite{2008Ghosh,2019Taffet,2020Valentine,2020Belov,2021Belov,DMRGSchollwock} due to the rapid exponential decay of the eigenspectrum of the density matrix. The iterative nature of the infinite-system algorithm enables the calculation and storage of the required wavefunctions at each growth step. As described in Section \ref{Se:2.4}, our definition of the triplet-pair population requires us to calculate the $N_d = N/2$ lowest members of the lowest energy family of triplet excitons, where $N$ is the number of C-atoms in the chain. For the final system size, we only calculate the $\ag{2}$ state. At each chain growth step one sweep of the chain was used to reduce the error and assure convergence.

\begin{figure}[b]
    \centering
    \includegraphics[width=0.7\linewidth]{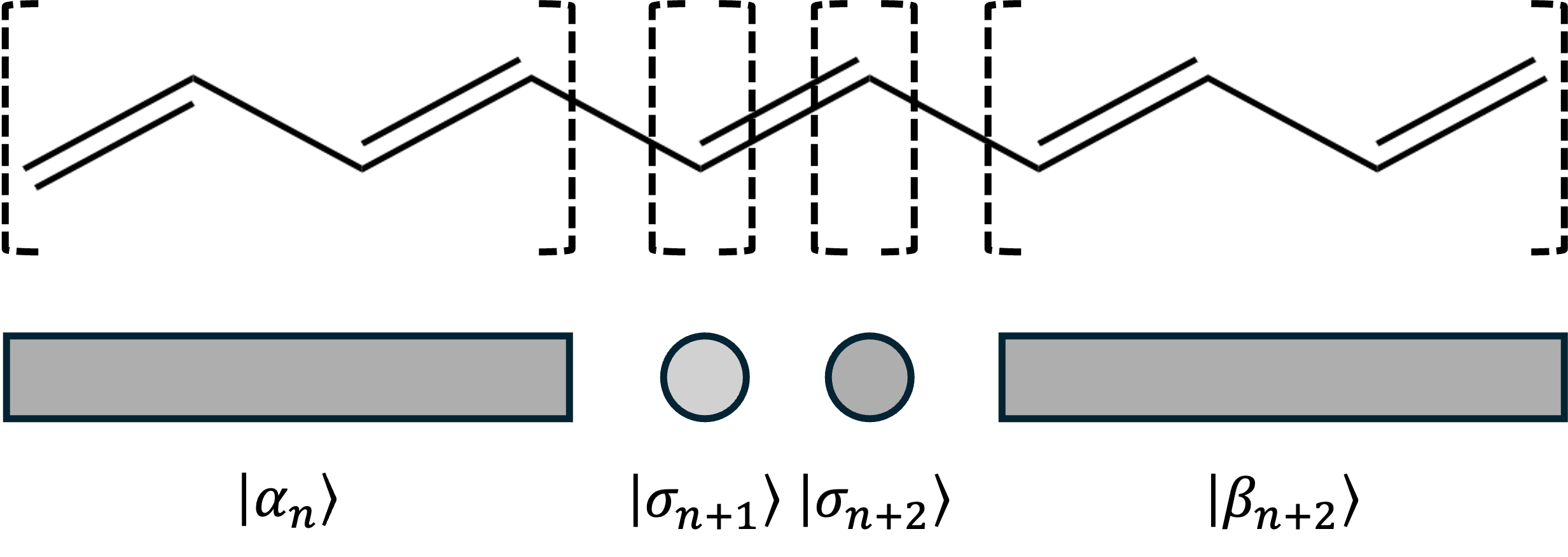}
    \caption{The structure of the DMRG blocks. The first and last blocks are described by the set of block states $\{|\alpha_n\rangle \}$ and $\{|\beta_{n+2}\rangle \}$ respectively, which correspond to linear combinations of tensor products of site states, $|\sigma_m\rangle$. Above the block diagram is a representation of the respective segments of a real chain that these block states describe. The site states describe the occupation of the carbon $p_z$ orbitals, i.e., $|\sigma\rangle =\{ |0\rangle, |\uparrow\rangle, |\downarrow\rangle,|\uparrow \downarrow \rangle\}$.}
    \label{fig:DMRG_block}
\end{figure}

The wavefunctions obtained from DMRG~\cite{1992White,1993White} are of the form
\begin{equation}
    |\Psi_{N}\rangle = \sum_{\alpha_n\sigma_{n+1}\sigma_{n+2}\beta_{n+2}} \Psi_{\alpha_n\sigma_{n+1}\sigma_{n+2}\beta_{n+2}}|\alpha_n\rangle |\sigma_{n+1}\rangle |\sigma_{n+2}\rangle |\beta_{n+2}\rangle,
\end{equation}
where the state $|\sigma_m\rangle$ describes the $m$th C-atom in a chain of $N$ C-atoms, and the block states $|\alpha_n\rangle$ and $|\beta_{n+2}\rangle$ describe the two DMRG blocks, as shown in Fig.~\ref{fig:DMRG_block}. We can view DMRG wavefunctions via the Matrix Product State (MPS) formalism~\cite{1995Ostlund,MPSSchollwock},  which allows us to write the block states as a linear combination of tensor products,
\begin{widetext}
\begin{equation}
    |\alpha_n\rangle = \sum_{\alpha_1,\alpha_2,\ldots,\alpha_{n-1}} \sum_{\sigma_1, \sigma_2,\ldots,\sigma_n}
    A_{\alpha_1}^{\sigma_1} A_{\alpha_1,\alpha_2}^{\sigma_2}\ldots A_{\alpha_{n-1},\alpha_{n}}^{\sigma_n} |\sigma_1 \sigma_2\ldots \sigma_n\rangle,
\end{equation}
\end{widetext}
where $A_{\alpha_{m-1},\alpha_m}^{\sigma_m}$ is the MPS tensor corresponding to the $m$th C-atom in the left block. Likewise, the states describing the right block are
\begin{widetext}
\begin{equation}
    |\beta_{n+2}\rangle = \sum_{\beta_{n+3},\ldots,\beta_{N-2},\beta_{N-1}} \sum_{\sigma_{n+3},\ldots,\sigma_{N-1}, \sigma_{N}}
    B_{\beta_{n+2},\beta_{n+3}}^{\sigma_{n+3}}\ldots B_{\beta_{N-2},\beta_{N-1}}^{\sigma_{N-1}}B_{\beta_{N-1}}^{\sigma_{N}} |\sigma_{n+3} \ldots \sigma_{N-1}\sigma_{N}\rangle,
\end{equation}
\end{widetext}
where $B_{\beta_{m-1},\beta_{m}}^{\sigma_m}$ is the MPS tensor corresponding to the $m$th C-atom in this block.

This form of the wavefunction is of better use to us, as it allows us to compute wavefunction overlaps straight-forwardly~\cite{MPSSchollwock}. We can always write all wavefunctions in the same (site) basis, irrespective of the truncation procedure or rotation of the block basis involved in DMRG.

\subsection{Two-particle representation of the triplet-pair state}\label{Se:2.3}


In this section we motivate our procedure to determine the triplet-pair population of the $\ag{2}$ state. We do this by the simpler problem of computing the two-particle population of a general many-particle eigenstate of the Hamiltonian $\hat{H}=\hat{H}_1 + \hat{H}_2$, where
\begin{equation}
    \hat{H}_{1}=-t\sum_{n=1}^{N_d-1}|n\rangle \langle n+1|+|n+1\rangle \langle n|,
\end{equation}
and
\begin{equation}
    \hat{H}_2= -V\sum_{n=1}^{N_d-1}|n;n+1\rangle \langle n;n+1|.
\end{equation}
$\hat{H}_1$ describes the hopping of  indistinguishable particles between neighboring sites along a linear chain of $N_d$ sites, while
$\hat{H}_2$ describes  a nearest-neighbor two-particle interaction. We assume a hard-core repulsion between the particles.

\begin{figure}
    \centering
    \includegraphics[width=0.7\linewidth]{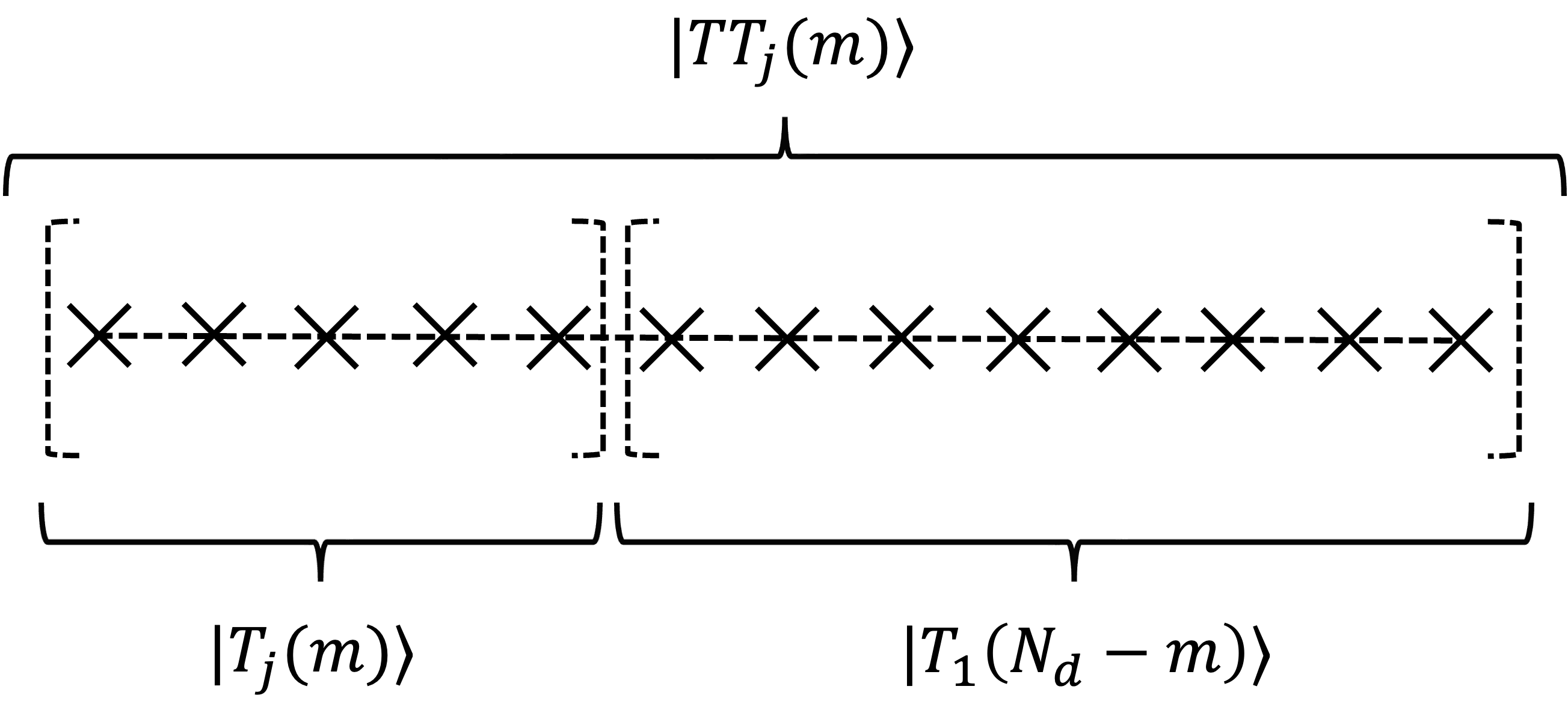}
    \caption{The definition of our two-particle (or triplet-pair) basis, described in Section \ref{Se:2.3} (and Section \ref{Se:2.4}). $\times$ represents a site (or dimer).
    For a chain of $N_d$ sites (or dimers), we define a two-particle (or triplet-pair) state as the tensor  product between a single-particle (or triplet) state $|T_j(m)\rangle$ occupying the left subchain of $m$ sites (or dimers)  and a single-particle (or triplet) state $|T_1(N_d-m)\rangle$ occupying the right subchain of $(N_d-m)$ sites (or dimers).
    $1 \le m \le (N_d - 1)$ and 
$1 \le j \le  m$, such that $\sum_{m=1}^{N_d-1} m = N_d(N_d-1)/2$.}
    \label{Fig:TT_pair_basis.png}
\end{figure}

We construct a  two-particle basis of size $N_d(N_d-1)/2$  for a chain of $N_d$ sites composed of two subchains of length $m$ and $(N_d - m)$  by a tensor product 
of the one-particle states, as illustrated in Fig.\ \ref{Fig:TT_pair_basis.png}. The tensor product is defined as
\begin{equation}
    |TT_j(m)\rangle =  |T_j(m)\rangle \otimes |T_1(N_d-m)\rangle,
\end{equation}
where
\begin{equation}\label{Eq:8}
    |T_j(m)\rangle = \left(\frac{2}{m+1}\right)^{1/2}\sum_{n=1}^m \sin\left(\frac{\pi jn}{m+1}\right)|n\rangle
\end{equation}
is the one-particle eigenstate of $\hat{H}_1$ for a subchain of $m$ sites.
The length of the left-hand subchain, $m$, satisfies $1 \le m \le (N_d - 1)$, while the quantum number $j$ satisfies
$1 \le j \le  m$, such that the basis $\{ |TT_j(m)\rangle \}$ spans the two-particle Hilbert space, i.e., 
$\sum_{m=1}^{N_d-1} m = N_d(N_d-1)/2$.

Next, the basis $\{ |TT_j(m)\rangle \}$ is orthonomalized via the L\"owdin symmetric orthogonalization~\cite{1950Lowdin} procedure to yield the basis $\{ |{TT}_j(m) \rangle\rangle \}$ (denoted by the double-ket).

We explicitly demonstrate numerically that this basis is a complete representation of any two-particle eigenstate, $|\Psi\rangle$, of $\hat{H}$ by computing the population,
\begin{equation}
  P =  \sum_{m=1}^{N_d-1}\sum_{j=1}^{m}\left |\langle\langle TT_j(m)|\Psi\rangle \right|^2.
\end{equation}
$P = 1$ for any choice of the interaction, $V$, or for any state $|\Psi\rangle$.

\subsection{Definition of the  triplet-pair population of the $\ag{2}$ state}\label{Se:2.4}

The mapping of the two-particle problem  described in Section \ref{Se:2.3} onto the calculation of the triplet-pair population of the $\ag{2}$ state 
is by the observation that the ket $|n\rangle$ in Eq.\ (\ref{Eq:8}) represents a triplet localized on the $n$th ethylene-dimer. In addition, $ |T_j(m)\rangle$ represents a triplet eigenstate on a subchain of $m$ dimers  and the two-particle state $|\Psi\rangle$ represents the $\ag{2}$ state of the full chain of $N_d$ dimers, both computed from the  PPP model  via DMRG. 
For a subchain of $m$ dimers $ |T_j(m)\rangle$ is selected from the $m$ members of the lowest-lying covalent triplet family, i.e., those with the same particle-hole symmetry as the $\ag{2}$ state.

We then use the MPS representation of the DMRG wavefunction to compute the tensor product
\begin{equation}
  |TT_j(m)\rangle =   |T_j(m)\rangle \otimes |T_1{(N_d-m)}\rangle,
\end{equation}
where  $1 \le m \le (N_d -1)$ and $1 \le j \le m$, again illustrated in Fig.~\ref{Fig:TT_pair_basis.png}. We next orthonormalize these triplet-pair states via the L\"owdin orthogonalization~\cite{1950Lowdin} to yield the basis  $\{ |TT_j(m)\rangle\rangle \}$ and again using the MPS formulation we compute the projection 
$\langle \langle TT_j(m)| 2^1A_g\rangle$.

The triplet-pair population, $P_{TT}$, of the $\ag{2}$ state is now defined as
\begin{equation}\label{TT_weight}
    P_{TT} = 3\sum_{m=1}^{N_d-1} \sum_{j=1}^{m} \left| \langle \langle TT_j(m)| 2^1A_g\rangle  \right|^2,
\end{equation}
where the factor of 3 appears because we only compute triplet-pairs where the triplets in both subchains have  a spin-projection, $S_z = 0$, i.e., the triplet-pairs are  $T_0 \otimes T_0$, where  the subscript refers to the $m_S$ eigenvalue of that triplet. A triplet-pair in an overall singlet state is 
\begin{equation}
|{^1}TT\rangle = \frac{1}{\sqrt{3}} \left( |T_{+1}T_{-1}\rangle - |T_{0}T_{0}\rangle + |T_{-1}T_{+1}\rangle  \right).
\end{equation}
Since the $2^1A_g$ state is a spin eigenstate with $S=0$ eigenvalue, it must also mean that any triplet-pair component must be found in an overall spin-symmetrised singlet state. It follows that the triplet-pair population $\left| \langle {^1}TT | 2^1A_g \rangle \right|^2 = \frac{1}{3}\left| 3 \langle T_0 T_0 | 2^1A_g \rangle \right|^2 = 3 \left| \langle T_0 T_0 | 2^1A_g \rangle \right|^2$, which is Eq.\ (\ref{TT_weight}).


In addition to the triplet-pair population, we also calculate the triplet-pair binding energy, $\Delta E_{TT}$, defined as
\begin{equation}\label{Eq:14}
    \Delta E_{TT} = E_{1^5A_g} - E_{2^1A_g}.
\end{equation}
We use the energy of the lowest quintet state $1^5A_g$ as a proxy for the energy of two electronically uncorrelated triplets. Based on an energetics argument, Valentine et al.~\cite{2020Valentine} suggested that the lowest quintet state in polyenes has predominantly unbound triplet-pair character. We calculate the triplet-pair population $\frac{3}{2}\sum_{m,j}\left| \langle \langle TT_j(m)|1^5A_g\rangle \right|^2$, of the lowest quintet state for chains of 8, 10, 12 and 14 C-atoms, i.e., $N_d =$ 4, 5, 6 and 7 ethylene dimers respectively) for all values of the Coulomb interaction parameter considered. This is further discussed in Appendix~\ref{App:A}. In all cases, $P_{TT} \gtrsim 90\%$, which is further evidence that the quintet state does indeed correspond to a pair of unbound triplets.

Notice that although the two-particle basis defined  in Section \ref{Se:2.3} is complete, it does not necessarily follow that our triplet-pair basis constructed from the DMRG calculation of the PPP model is also complete. Therefore, our calculated triplet-pair population is formally a lower bound. This is confirmed by our estimate of the triplet-pair population of the lowest energy quintet state of the PPP model as being only ca.\ 93\%, rather than 100\%. However, as noted by previous authors~\cite{1987Tavan,2019Taffet, 2020Valentine}, the $|T_1 \otimes T_1 \rangle$ triplet-pairs are expected to form the largest contribution to the $2^1A_g$ state. Our triplet-pair basis accounts for every partition of a chain into two $T_1$ triplet states, with smaller contributions from higher energy triplet-pairs. Therefore, the triplet-pair population of the $2^1A_g$ state as calculated in this paper is expected to account for almost all of the triplet-pair character of the dark state.

\section{Results}\label{Se:3}

\begin{figure}[t!]
    \centering
    \includegraphics[width=0.7\linewidth]{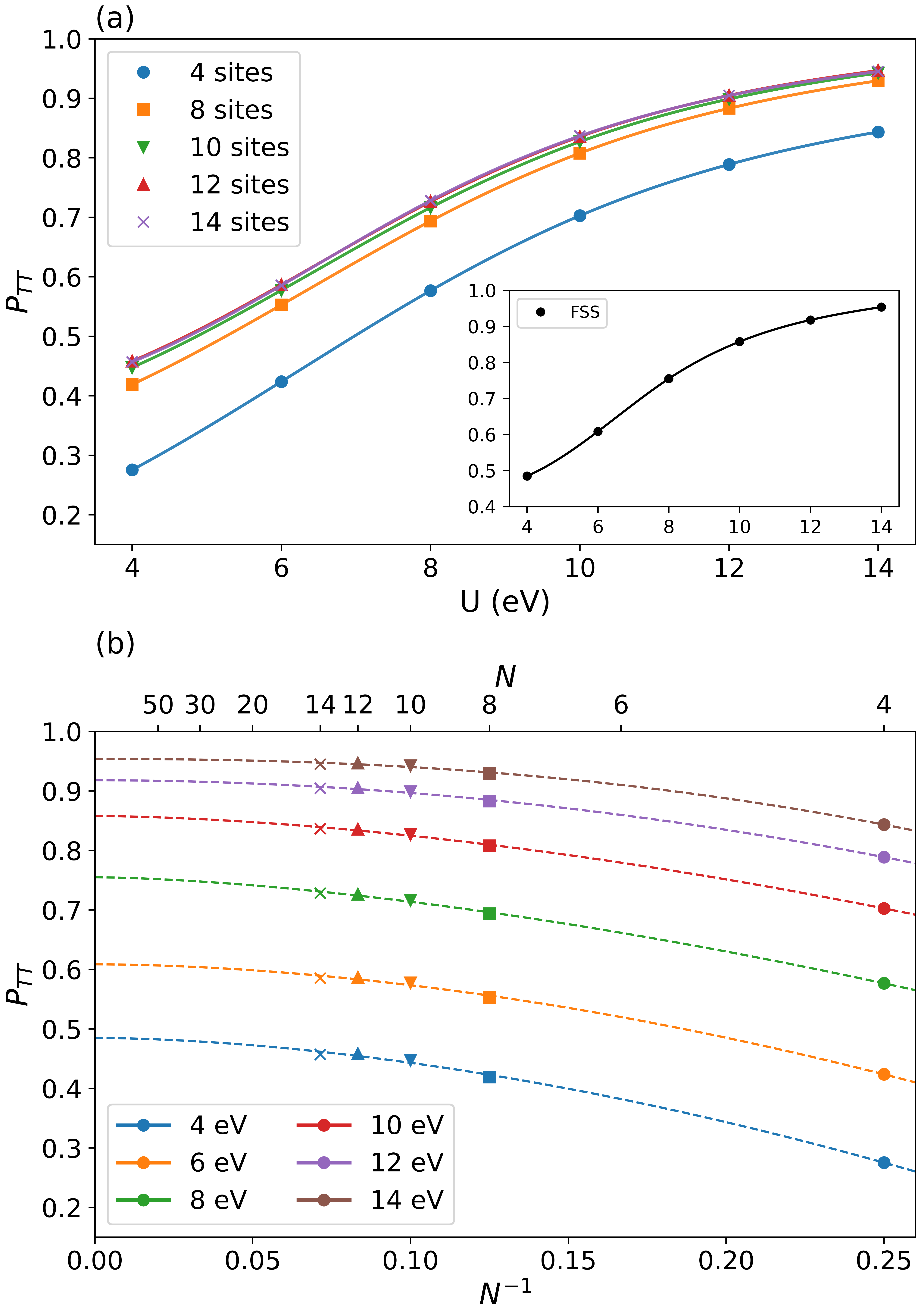}
    \caption{(a) The triplet-pair population, $P_{TT}$, of the $2^1A_g$ state (defined in Eq.\ (\ref{TT_weight})) as a function of the Coulomb interaction parameter, $U$. (b) The extrapolation of the finite-size-scaling (FSS) values of $P_{TT}$ for the six different Coulomb parameters versus  $N^{-1}$, where $N$ is the number of C-atoms. The data was fitted to the function $P_{TT} = aN^{-\alpha} + c$. The inset in (a) shows the FSS values of $P_{TT}$ as a function of $U$. Also shown are the exact values of $P_{TT}$ for a 4 C-atom chain, as described in Appendix~\ref{AppB:Huckel}.}
    \label{fig:W_TT_vs_U}
\end{figure}

The calculated values of $P_{TT}$ as a function of the Coulomb parameter, $U$, are shown in Fig.~\ref{fig:W_TT_vs_U}(a). The triplet-pair population increases with increasing Coulomb interaction, reaching the asymptotic limit of $P_{TT} \rightarrow 1$ as $U/t_0 \rightarrow \infty$ for all chain lengths. That the $\ag{2}$ state becomes entirely triplet-pair in character in the strong-coupling limit is a known result~\cite{1999Chandross, 2022Barford, WB_Book}. In this limit all determinants corresponding to each $p_z$ orbital that are singly occupied by an electron are degenerate~\cite{1976Schulten}, meaning that there are $2^{N}$ degenerate eigenstates with the same energy as the $1^1A_g$ state, including the $2^1A_g$ state. The $\ag{2}$ state corresponds to two electron spin flips with respect to $1^1A_g$~\cite{1987Tavan, WB_Book}. Since there are no charge-transfer components,
this implies that the $\ag{2}$ state is entirely  triplet-pair  in character. 

On the other hand, we  note that the triplet-pair population does not vanish in the opposite, noninteracting limit, as one might expect from the molecular orbital description.
As shown in Appendix~\ref{AppB:Huckel} and Ref~\cite{1999Chandross}, contributions to the ground state from charge-transfer states between ethylene dimers result a in nonzero triplet-pair character in the $2^1A_g$ state. This is a consequence of using the real-space valence bond picture to analyse the noninteracting eigenstates. Within the molecular orbital (or band) picture the $2^1A_g$ state is a single electron-hole (e-h) excitation, i.e., the $\ag{2}$ state does not have any doubly-excited e-h character. Furthermore, the increase in the triplet-pair character with increasing Coulomb interaction is a consequence of the covalent states (i.e., triplet-pairs) becoming more energetically favorable with respect to the ionic contributions (i.e., charge-transfer excitons) in the $\ag{2}$ state~\cite{2022Barford,WB_Book,1999Chandross}.

\begin{figure}[t]
    \centering
    \includegraphics[width=0.7\linewidth]{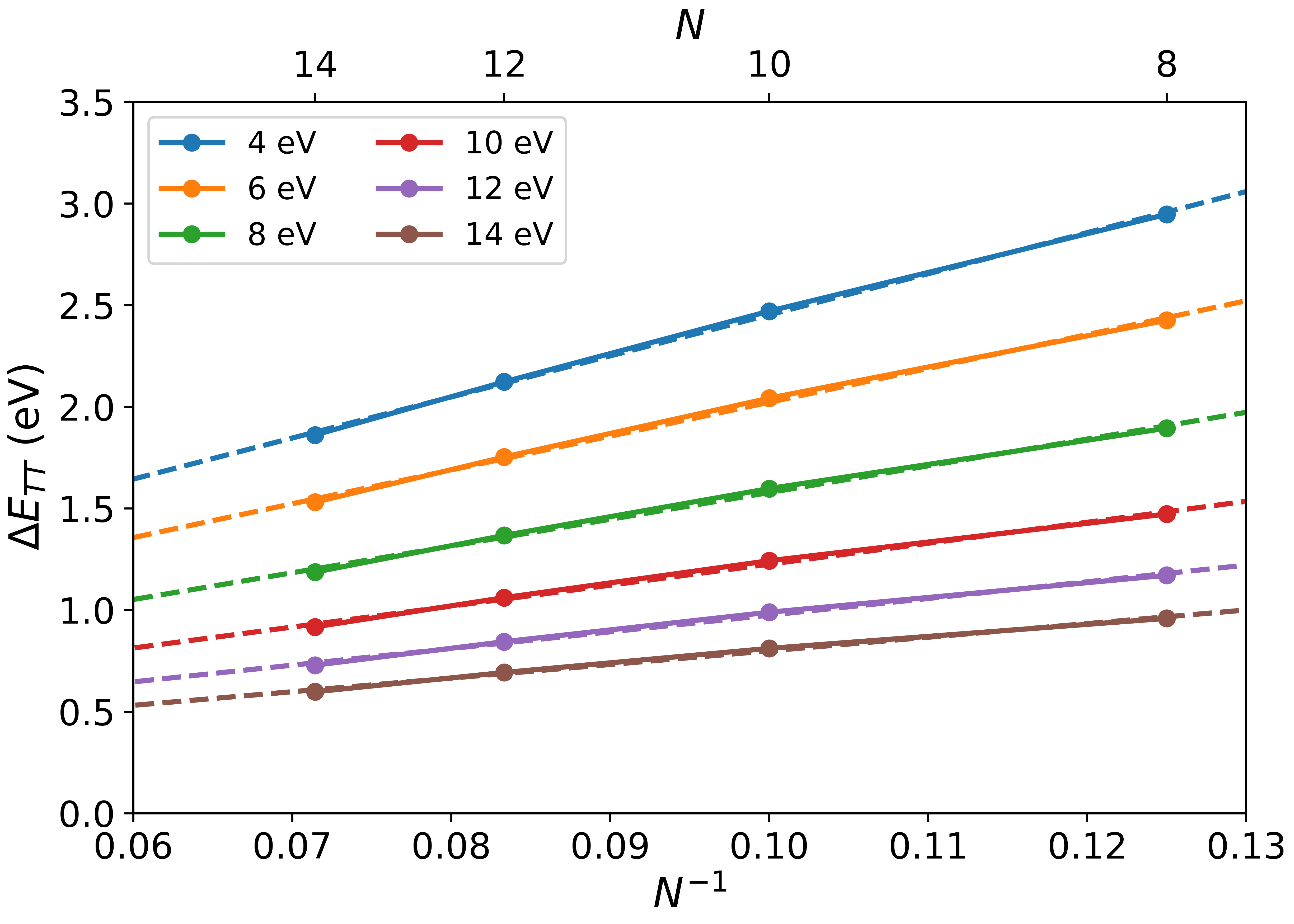}
    \caption{The triplet-pair binding energy, $\Delta E_{TT}$, (defined in Eq.\ (\ref{Eq:14}))  as a function of $N^{-1}$, where $N$ is the number of C-atoms, for different values of the Coulomb parameter, $U$.}
    \label{fig:dE_vs_N}
\end{figure}

The dependence of $P_{TT}$ on chain length is shown in Fig.~\ref{fig:W_TT_vs_U}(b). We notice an increase in $P_{TT}$ with increasing number of C-atoms, $N$, but no obvious relationship emerges between the triplet-pair population and the size of the system from our data. We fitted our data to the power function $P_{TT}(N)=aN^{-\alpha}+c$. We see that this predicts a monotonic increase in the triplet-pair population of the $\ag{2}$ state with increasing number of C-atoms. This behavior correlates with the change in the binding energy, shown in Fig.~\ref{fig:dE_vs_N}. The triplet-pair binding energy and the triplet-pair population are related quantities: $P_{TT}$ increases with decreasing $\Delta E_{TT}$~\cite{2022Barford}.
The finite-size-scaling (FSS) result for the triplet-pair population as a function of Coulomb interaction is shown in the inset of Fig.~\ref{fig:W_TT_vs_U}(a). The realistic value of $U$ for $\pi$-conjugated systems~\cite{1997Chandross} is $U = 8$ eV, and thus we deduce that the triplet-pair population, $P_{TT}$, in polyenes is ca.\ 75\%. Note that this value, although remarkably close to the calculated values of the doubly-excited character of the $2^1A_g$ state \cite{2016Dreuw,2019Boguslawski}, does not necessarily imply that the $2^1A_g$ state has 75\% doubly excited character, as shown in Appendix~\ref{AppB:Huckel} for the noninteracting limit. The triplet-pair population is a quantity closely related to the doubly-excited character (i.e., the weight of the doubly-excited determinants) of a wavefunction but not equivalent to it, as one is defined within real space and the other within the molecular orbital theory. However, both are seen to follow similar trends~\cite{2016Dreuw,2019Boguslawski,2023Plasser,2025Plasser}.

\vfill\pagebreak

\section{Discussion and conclusions}\label{Se:4}

In conclusion, we have presented the results of numerical calculations of the real-space triplet-pair population of polyenes of 8, 10, 12, and 14 C-atoms for different values of the Coulomb interaction. Using these results and the correlation between the triplet-pair binding energy and triplet-pair character of the $\ag{2}$ state, we extrapolated the value of the triplet-pair population to infinitely long chains. Our calculations agree with other measures of double-excitation character derived from ab initio calculations~\cite{2016Dreuw,2019Boguslawski,2023Plasser,2025Plasser}, where an increase in the triplet-pair character with increasing chain size is observed. Furthermore, we notice that the triplet-pair population also agrees with calculations of Chandross et al.~\cite{1999Chandross} in the limit of noninteracting electrons where $P_{TT}$ is small but nonzero, as well as in the strongly-interacting limit where the $\ag{2}$ state $P_{TT} \rightarrow 1$.

Our results have implications for possible singlet-fission mechanisms in carotenoids that postulate that one of  the $\ag{2}$-family~\cite{2020Valentine} is the intermediary between the bright-excited state and spatially uncorrelated, decohered triplet-pairs. However, since the lowest-energy member of this family (i.e., the $\ag{2}$ state) is composed of a strongly bound triplet-pair~\cite{2020Valentine,2022Barford}, potential singlet fission from this state would be endothermic, whereas potential singlet fission from higher-energy members would be exothermic~\cite{2020Valentine,2023Barford}. The triplet-pair population of these higher-energy members of the $\ag{2}$-family are therefore discussed in Appendix \ref{App:A}.

\begin{acknowledgments}
A.G.I. thanks Linacre College for the Carolyn and Franco Gianturco Scholarship.
\end{acknowledgments}

\appendix

\section{Triplet-pair population of the $2A_g$-family of states}\label{App:A}

 The $\agp{2}$ state is the lowest energy member (with the lowest pseudomomentum) of a band of states that corresponds to the  triplet-pair quasiparticle~\cite{2020Valentine}. The higher-energy (higher pseudomomentum) states are $\bup{1}, \agp{3}, \cdots$. Here, we explicitly label the states by their particle-hole eigenvalue of $+1$ so as to distinguish the $\bup{1}$ state from the optically bright $\bum{1}$ state \footnote{Note that in the quantum chemistry literature, the $2A_g$-family are labeled 
 $\agm{2}, \bum{1}, \agm{3}, \cdots$.}.

 \begin{figure}
    \centering
    \includegraphics[width=0.7\linewidth]{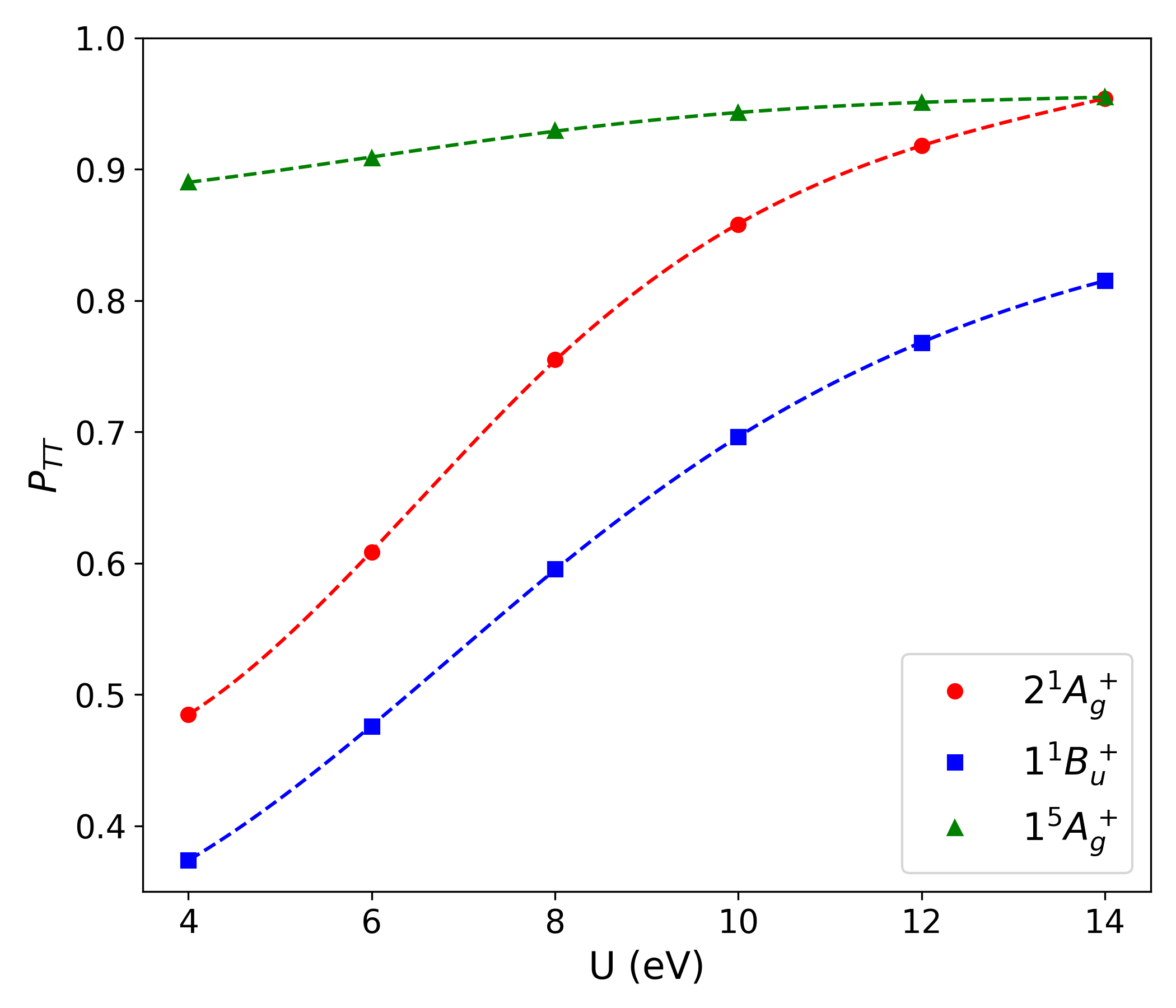}
    \caption{The triplet-pair population of the first and second lowest energy dark states, $2^1A_g^+$ and $1^1B_u^+$, in the limit of infinite chains. Also shown is the triplet-pair population calculated for the lowest energy quintet state, $1^5A_g^+$.}
    \label{fig:appA}
\end{figure}

As discussed in Section \ref{Se:4}, although the $\agp{2}$ state has a large triplet-pair population,  singlet-fission from this state would be endothermic, owing to the strong triplet-pair attraction~\cite{2020Valentine,2022Barford}.
In contrast, triplets in the  higher-energy members of  the $\ag{2}$-family are weakly bound and intermolecular  singlet-fission from these states would be exothermic~\cite{2020Valentine,2023Barford}.  Thus, the triplet-pair population of these states is relevant for potential singlet-fission mechanisms in polyenes.

Fig.~\ref{fig:appA}  shows the triplet-pair population of the $2^1A_g^+$ and $1^1B_u^+$ states, in the limit of infinite chains (as predicted in the same way as described in the main text). The triplet-pair population is observed to decrease as the pseudomomentum increases, i.e., higher energy dark states have a smaller $P_{TT}$. This can be rationalized by the coupling between the singlet triplet-pair subspace and the higher energy singlet charge-transfer exciton subspace, as described  in Ref.~\cite{2022Barford}. This coupling results in a series of eigenstates, where the lowest energy eigenstate has the highest triplet-pair weight. Furthermore, our result agrees with the ab initio prediction in Ref.~\cite{2016Dreuw} that the doubly excited character of the $1^1B_u^+$ state is smaller than that of $2^1A_g^+$ state. (The spin correlations and charge-transfer exciton wavefunctions of the $\agp{2}$, $\bup{1}$ and $\agp{3}$ members of the $\ag{2}$-family are illustrated in Fig.\ 7 and Fig.\ 8 of Ref.~\cite{2020Valentine}, respectively.)


The extrapolated triplet-pair population of the lowest quintet state, $1^5A_g^+$, is also shown in Fig.~\ref{fig:appA}. For a chain of 4 C-atoms, a quintet state can only be formed by the coupling of two triplet states in an overall $S=2$ state. Therefore, for butadiene, this quintet state has $P_{TT} = 1$. Applying our definition of $P_{TT}$ for chains of 8 to 14 C-atoms we find $0.90 \le P_{TT} \le 0.96$, where the triplet-pair population increases with increasing Coulomb parameter and decreases with increasing chain length. 


\section{Triplet-pair population in the noninteracting limit}\label{AppB:Huckel}

In this appendix we demonstrate that in the noninteracting limit in a chain of four C-atoms (i.e., sites)  the $\ag{2}$ state has a nonzero triplet-pair population.

The noninteracting $\pi$-electron Hamiltonian is
\begin{equation}
    \hat{H} = -\sum_{i,\sigma }t_i(\hat{c}_{i\sigma}^\dagger \hat{c}_{i+1\sigma} + \hat{c}_{i+1\sigma}^\dagger \hat{c}_{i\sigma}),
\end{equation}
where $t_i =t_0(1 +(-1)^{(i+1)} \delta)$ and we take $\delta = 1/12$.

On a chain of four-sites the $2^1A_g$  state  written in the site basis is
\begin{widetext}
\begin{equation}\label{Eq:B2}
    |2^1A_g\rangle = 0.1794 \left( |\uparrow \downarrow \uparrow \downarrow \rangle + |\downarrow \uparrow \downarrow \uparrow \rangle - |\downarrow \downarrow \uparrow \uparrow \rangle - | \uparrow \uparrow \downarrow \downarrow\rangle \right) + \ldots,
\end{equation}
\end{widetext}
where we only show the covalent (i.e., singly occupied) contributions to the  state.

Partitioning the four-site chain into two subchains of two sites each, we can write the triplet states of each subchain (or dimer) as
\begin{equation}
    |T_{+1}\rangle = |\uparrow \uparrow \rangle,
\end{equation}
\begin{equation}
    |T_0\rangle =  \frac{1}{\sqrt{2}}\left( |\uparrow \downarrow \ \rangle + |\downarrow \uparrow  \rangle \right)
\end{equation}
and
\begin{equation}
    |T_{-1}\rangle =  | \downarrow  \downarrow \rangle,
\end{equation}
for the $M_s = +1, 0$ and $-1$ spin-projections, respectively.

The triplet-pair basis for the four-site (two-dimer) chain that spans the $S_z = 0$ subspace is
\begin{equation}\label{Eq:B6}
    |T_{+1}T_{-1}\rangle = |\uparrow \uparrow \downarrow \downarrow \rangle,
\end{equation}
\begin{equation}\label{Eq:B7}
    |T_0 T_0\rangle =  \frac{1}{2}\left( |\uparrow \downarrow \uparrow \downarrow \ \rangle + |\uparrow \downarrow \downarrow \uparrow  \rangle +
  |\downarrow \uparrow \uparrow \downarrow \rangle  + |\downarrow \uparrow \downarrow \uparrow \rangle\right)\\
\end{equation}
and
\begin{equation}\label{Eq:B8}
    |T_{-1}T_{+1}\rangle = |\downarrow \downarrow  \uparrow \uparrow \rangle.
\end{equation}
The projection of each of these triplet-pair basis  states onto the $\ag{2}$ state (given in Eq.\ (\ref{Eq:B2})) is $0.1794$ and thus the total triplet-pair population of this state is 9.66\%.

Alternatively, using Eq.\ (\ref{Eq:B6}), Eq.\ (\ref{Eq:B7}) and  Eq.\ (\ref{Eq:B8}), we can construct the singlet triplet-pair state for the four-site chain as
\begin{equation}
|{^1}TT\rangle = \frac{1}{\sqrt{3}} \left( |T_{+1}T_{-1}\rangle - |T_{0}T_{0}\rangle + |T_{-1}T_{+1}\rangle  \right).
\end{equation}
Projecting this state onto the $\ag{2}$ state, the triplet-pair population is $\left| \langle {^1}TT|2^1A_g\rangle \right|^2 = 9.66\%$,
as before. This value is the exact triplet-pair population because, for a chain of four C-atoms, $|^1TT\rangle$ spans the whole $S=0$ space of electron configurations that can be written as a product of two triplet states. 

The results of the same analysis for the values of the Coulomb parameter considered in the main text are shown in Fig.~\ref{fig:W_TT_vs_U}, which shows that the triplet-pair population increases as the Coulomb interaction increases.



\bibliography{2Ag_paper}

\end{document}